# The Sample Pre-selection and Characterization Station at the SECUF: Instrumentation, Capabilities, and Representative Scientific Achievements

Xu Chen(陈旭)[1*], Tao Sun(孙涛)[1], Huifen Ren(任会芬)[1], Minjie Cui(崔敏杰)[1], Jun Luo(罗军)[1], Shuai Zhang(张帅)[1] and Shaokui Su(苏少奎)[1*]

[1]*Beijing National Laboratory for Condensed Matter Physics, Institute of Physics, Chinese Academy of Sciences, Beijing 100190, China*

**Abstract:** The Synergetic Extreme Condition User Facility (SECUF) is a comprehensive, state-of-the-art user facility designed to provide integrated extreme physical conditions—including ultrahigh pressure, ultralow temperature, strong magnetic fields, and ultrafast optical fields—for frontier research in condensed matter physics and materials science. Within SECUF, the F2 Sample Pre-selection and Characterization Station plays a pivotal supporting role. Its mission is to provide comprehensive sample synthesis, processing, pre-screening, and characterization services to prepare high-quality specimens for subsequent experiments under extreme conditions. This paper details the specifications and performance of ten core instrument systems within these units. Furthermore, we highlight several breakthrough scientific achievements enabled by the F2 Station, encompassing the discovery of novel quantum spin supersolid states, pressure-induced high-temperature superconductivity in nickelates, giant anomalous Hall angles, and molecular water in lunar soil. We also outline ongoing technical developments that expand the station's capabilities, such as integrated high-pressure cells and self-built ancillary measurement systems.

* To whom correspondence should be addressed: xchen@iphy.ac.cn sski@iphy.ac.cn

## 1. Introduction

The Synergetic Extreme Condition User Facility (SECUF) is an internationally advanced user facility that integrates multiple extreme physical conditions, including ultrahigh pressure, ultralow temperature, strong magnetic fields, and ultrafast optical fields. Its primary objective is to expand the research horizon of materials science, promote the discovery of novel states of matter, new phenomena, and new physical laws, and support China in achieving world-class research outcomes in condensed matter physics and related interdisciplinary fields. Key targeted research directions include the discovery of new high-temperature superconductors, breakthroughs in understanding unconventional superconducting mechanisms, advances in core quantum computing technologies, and ultrafast control of material properties.

The F2 Sample Pre-selection and Characterization Station is a critical auxiliary support platform within SECUF. It undertakes the essential tasks of sample pre-selection and characterization, specifically involving high-quality crystal growth and preparation, material composition and structural analysis, and the characterization of basic physical properties. The final goal is to screen and identify high-quality samples with exceptional properties for the next stage of investigation under even more extreme conditions.

The station is operationally organized into four specialized testing units: the Physical Property Measurement Unit, the Structure and Morphology Screening Unit, the Spectroscopic Characterization Unit, and the High-Quality Crystal Growth Unit. This configuration encompasses ten major instrument systems and dozens of auxiliary mid- to small-scale pieces of equipment. These units operate in a coordinated, well-defined, and synergistic manner, establishing a fully integrated, streamlined workflow from initial sample preparation to final advanced characterization. This integrated approach has proven highly effective in supporting user-led research. Users from various institutions, leveraging the advanced research infrastructure of the F2 station and supported by its instrument specialists, have achieved multiple outstanding scientific outcomes and made significant breakthrough progress.

## 2. Instrumentation and Technical Capabilities of the F2 Station

This section details the core instrument systems within the four operational units of the F2 Station, outlining their key specifications and recently developed technical extensions.

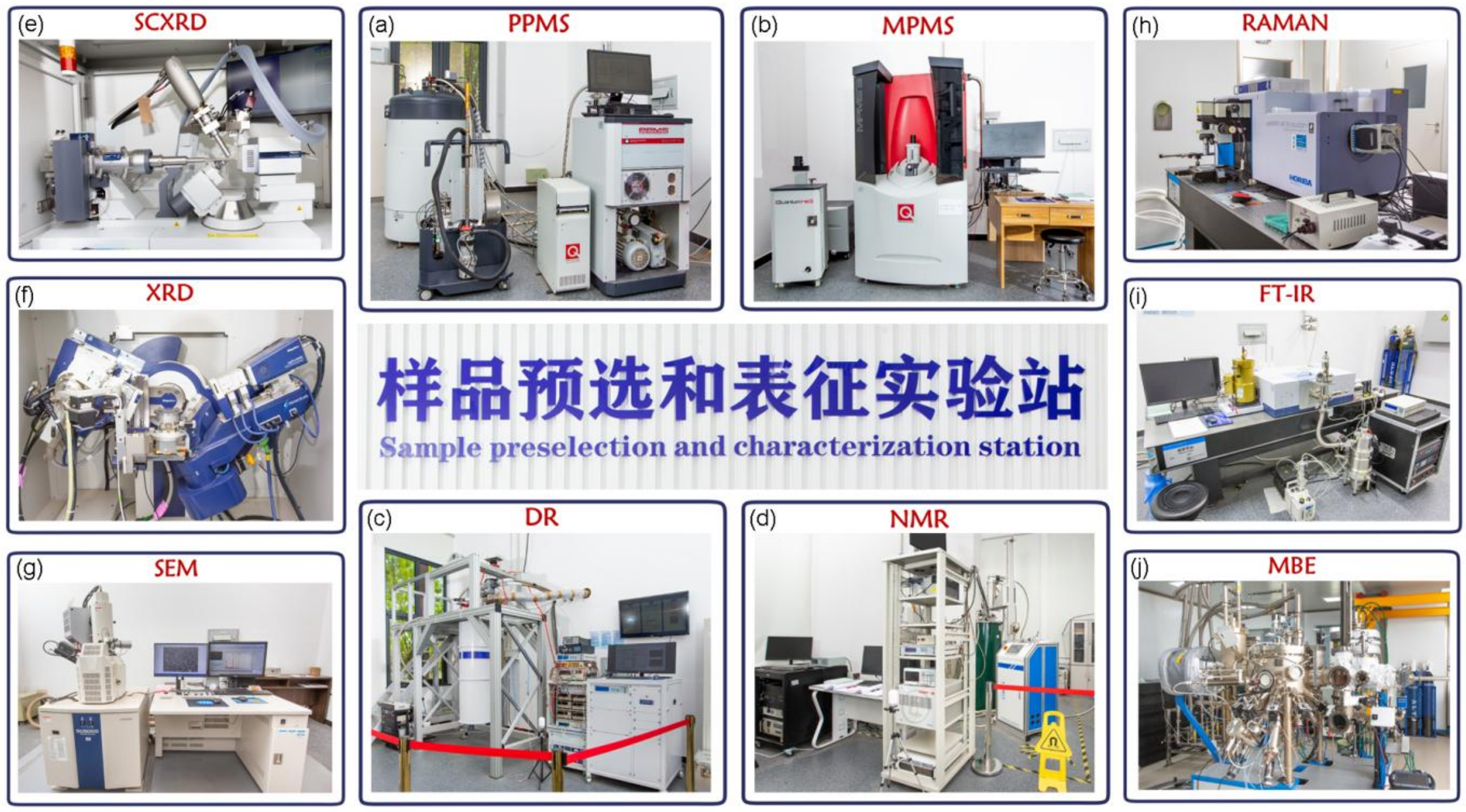

**Fig. 1. Instrumentation of the F2 station**, including (a) 16 T multifunctional ultra-low temperature physical property measurement system (PPMS), (b) Ultra-low temperature magnetic measurement system (MPMS), (c) Triton cryogen-free He3-He4 dilution refrigerator (DR), (d) 9 T nuclear magnetic resonance spectrometer system (NMR), (e) Single crystal X-ray diffractometer (SCXRD), (f) X-ray diffractometer (XRD), (g) Scanning electron microscope (SEM), (h) Micro confocal Raman spectrometer (Raman), (i) Fourier transform infrared spectrometer (FT-IR), (j) Molecular Beam Epitaxy system (MBE).

### 2.1 Physical Property Measurement Unit

#### 2.1.1. Quantum Design 16T-PPMS Multifunctional Measurement System (see Fig. 1(a))

The 16T Physical Property Measurement System (PPMS-16) provides a controlled low-temperature and high magnetic field environment for comprehensive electrical, thermal, and magnetic characterization. The system operates over a temperature range of 1.8 – 400 K (extendable down to 0.05 – 4 K with an optional Dilution Refrigerator insert), with a temperature sweep rate of 0.01 – 8 K/min and excellent stability (±0.2% for T < 10 K, ±0.02% for T > 10 K). It generates magnetic fields up to ±16 T, with a sweep rate of 0.1 – 100

Oe/s, a stability of 1 ppm/hour, and a low remnant field of 5 Oe. Available measurement options include Resistivity, Electrical Transport (ETO), Heat Capacity (HC), Vibrating Sample Magnetometry (VSM), AC Magnetic Susceptibility (ACMS II), and a Horizontal Rotator.

The station has independently developed a DAC-based in-situ high-pressure electrical measurement system that provides an integrated, one-stop workflow encompassing sample pressing, hole drilling, sample loading, pressurization, pressure calibration, and electrical transport testing, enabling combined ultrahigh pressure (up to 70 GPa), low temperature (down to 1.8 K), and high magnetic field (up to 16 T) testing.

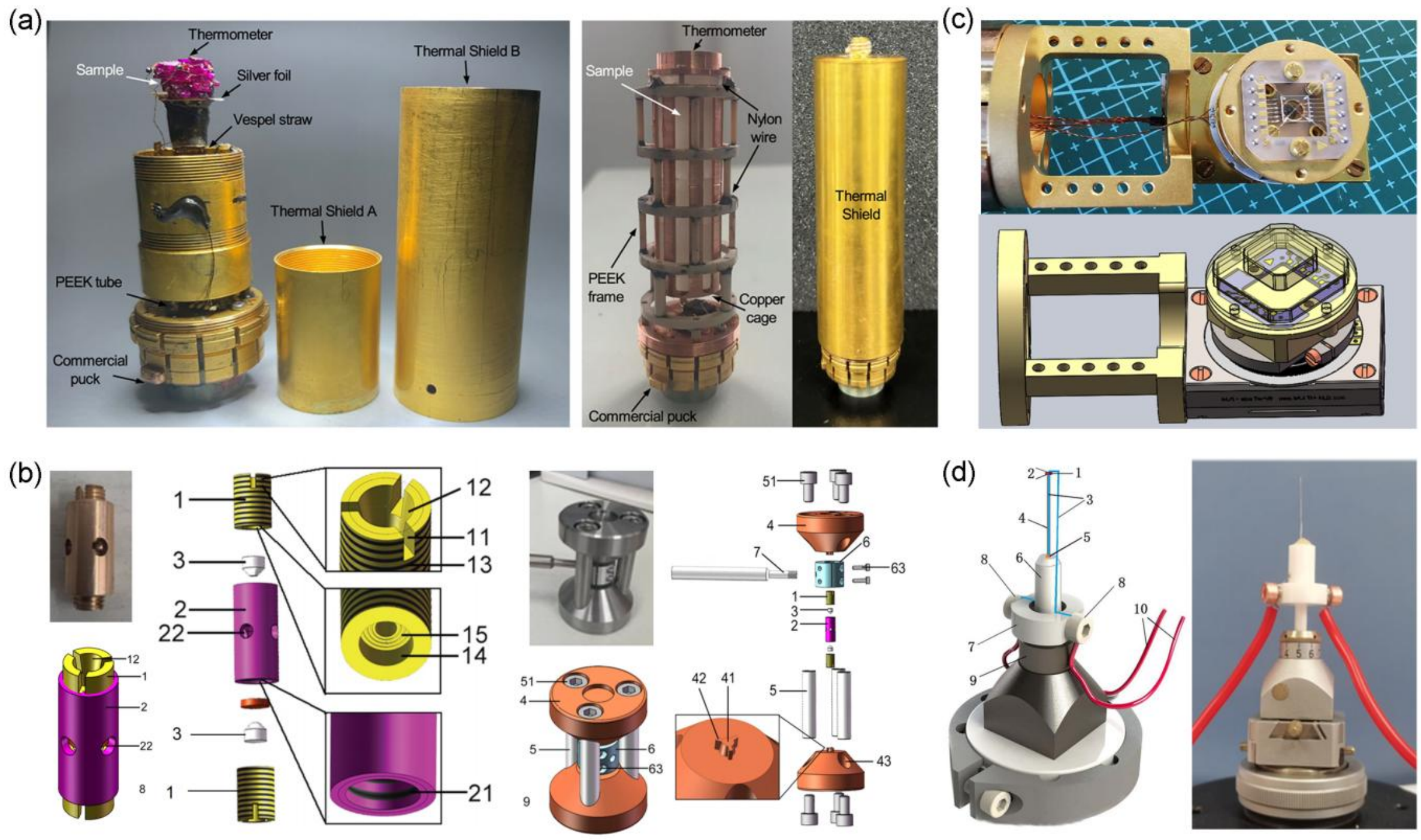

**Fig. 2. Instrument upgrade of the F2 station**, including (a) Photos of PPMS-based quasi-adiabatic demagnetization cooling measurement setups,[1] (b) Photos and diagrams of MPMS He3 option-based ultra-compact diamond anvil cell,[2,3] (c) Photos and diagrams of DR-based corner thermal sample holder, (d) Diagrams and photos of Single Crystal X-ray-based electric field controlled sample stage.[4]

Recently, the F2 station assisted users in developing a low-temperature quasi-adiabatic magnetocaloric effect (MCE) measurement setup, as shown in Fig. 2(a). Using $Na_2BaCo(PO_4)_2$ as the magnetocaloric material, adiabatic demagnetization was performed starting from the 2 K base temperature provided by the PPMS. This setup achieved a temperature reduction by more than an order of magnitude, ultimately reaching an ultra-low temperature below 90 mK.

This in-house development represents a significant technical breakthrough, as it extended the standard PPMS functionality to enable direct adiabatic demagnetization refrigeration measurements, and was instrumental in the discovery of the spin supersolid state and its giant MCE described in Section 3.1.

**2.1.2. Quantum Design MPMS3 Ultra-Low Temperature Magnetic Measurement System (see Fig. 1(b))**

The MPMS3 system is dedicated to ultra-sensitive magnetic measurements using both SQUID-VSM and DC measurement modes. It covers a temperature range of 1.8–400 K (extendable to 0.42–1.8 K with a $He^3$ option), with sweep rates of 10 K/min (T < 10 K) and 30 K/min (T > 10 K), and temperature stability specifications matching the PPMS. It provides magnetic fields up to ±7 T, sweepable at 2–700 Oe/s, with a low stray field (<0.01 G at 4 cm) and a remnant field of 5 Oe. The VSM offers a vibrating range of 0.1–8 mm, a high sensitivity of $<1\times10^{-8}$ emu at 0 T (for a 10-second measurement), and a maximum measurable moment of 10 emu within a 9 mm sample space.

The station has recently incorporated a Mcell high-pressure accessory (capable of reaching 20 GPa under routine operation, with a theoretical maximum of up to 60 GPa). When integrated with the MPMS3, this accessory enables sensitive high-pressure magnetic measurements on strongly magnetic materials, including superconductors, ferromagnets, and antiferromagnets, thereby substantially expanding the station's capabilities for magnetic characterization under high-pressure conditions. This combined high-pressure magnetic measurement capability is particularly valuable for studying pressure-induced quantum phase transitions in magnetic and superconducting systems.

Recently, based on a counter-rotating double-screw compression concept, we designed and fabricated an ultra-compact diamond anvil cell that overcomes the stringent size (diameter < 6 mm) and weight (< 3 g) constraints of the MPMS $He^3$ option, as shown in Fig. 2(b). This device features a simple structure and convenient operation, enabling DC magnetic susceptibility measurements down to the $He^3$ temperature range (0.4 K) and up to ultrahigh pressures of 20 GPa.[2,3]

### 2.1.3. Triton Cryogen-free Dilution Refrigerator System (see Fig. 1(c))

The Triton system integrates a $He^3$-$He^4$ Dilution Refrigerator with a 14 T superconducting magnet, providing a unique combined environment of ultra-low temperature (base temperature < 10 mK) and high magnetic field. The magnet can be ramped at rates below 0.3 T/min. The refrigerator provides cooling powers of 12 μW at 20 mK and 450 μW at 100 mK. Sample exchange is performed via a top-loading method into a sample space of < Ø30 mm. Standard measurement capabilities include AC/DC electrical resistivity and heat capacity.

We have also recently constructed an angular-dependent heat capacity measurement system specifically adapted for the dilution refrigerator. This system enables continuous angle-resolved heat capacity measurements under a fixed magnetic field at temperatures down to approximately 100 mK, achieving a measurement precision of 1 μJ/K, as shown in Fig 2(c). This capability is essential for investigating anisotropic superconducting gap structures and field-angle-resolved quantum oscillations in heavy-fermion and unconventional superconducting materials.

### 2.1.4. 9T Nuclear Magnetic Resonance Spectrometer (see Fig. 1(d))

This solid-state NMR system is centered on a 9 T superconducting magnet with homogeneity $< 3\times10^{-6}$ over a 10 mm diameter spherical volume, a maximum energization rate of 0.0088 T/s, and a stray field of 4.7 Gauss at 1.7 m. A $^4$He variable temperature insert controls the sample temperature from 300 K down to 1.9 K within a 40 mm sample space. The spectrometer operates over a frequency range of 1–400 MHz and utilizes a homemade NMR probe with cryogenic coaxial cables.

Recently, we integrated a piston-cylinder-type high-pressure cell into the system, thereby enabling NMR measurements under hydrostatic pressures up to 2.5 GPa. This extension opens the door for high-pressure NMR studies of quantum phase transitions and magnetic ordering in strongly correlated electron systems.

## 2.2 Structure and Morphology Screening Unit

### 2.2.1. Rigaku XtaLAB Synergy Single Crystal X-ray Diffractometer (see Fig. 1(e))

The XtaLAB Synergy is a full-featured diffractometer for both small molecule and protein crystallography. Its key feature is two independent X-ray sources (Cu and Mo Kα), allowing flexible wavelength selection. It is powered by a 1200W high-brilliance rotating anode generator (20–60 kV, 10–30 mA) and incorporates a high-precision 4-circle kappa goniometer and a HyPix-6000 detector. An optional liquid nitrogen cryostream enables measurements from 80 K to 400 K with ±0.1 K accuracy. The instrument performs indexing, lattice parameter determination, data collection, and structural analysis.

In collaboration with users, we have recently constructed an in-situ voltage-biased single-crystal diffraction sample stage based on the SCXRD instrument, enabling the structural characterization of materials under applied electric fields, as shown in Fig. 2(d).[4] Furthermore, we have introduced advanced electron-density refinement techniques that exploit high-quality single-crystal XRD data to determine the true electronic structure of light elements (e.g., hydrogen, boron, carbon, and oxygen). This approach goes beyond the conventional independent-atom model (IAM) by employing multipole refinement or maximum entropy methods, thereby enabling the direct visualization of valence electron distributions and chemical bonding features.[5]

#### 2.2.2. Rigaku SmartLab High-Resolution X-ray Diffractometer (see Fig. 1(f))

The SmartLab is a versatile high-resolution diffractometer for powder, bulk, and thin-film samples, controlled by the integrated SmartLab Studio II software. It uses a 9kW high-frequency X-ray generator with a Cu rotating anode (≤45 kV, ≤200 mA), a wide-angle θ-θ goniometer (-10° to 160°), and a 2D HyPix-3000 detector supporting 0D, 1D, and 2D measurement modes. Available options enhance capabilities for micro-area measurement, high-precision film analysis (with a tilt stage), in-plane geometry, and mapping. Supported techniques include XRD, glancing incidence (GI), X-ray reflectivity (XRR), rocking curve, reciprocal space mapping (RSM), in-plane diffraction, and micro-area analysis.

#### 2.2.3. Hitachi SU5000 Thermal Field Emission Scanning Electron Microscope (see Fig. 1(g))

The SU5000 SEM is equipped for comprehensive microanalysis with a high-vacuum secondary electron detector, an optional high-sensitivity BSE detector, and a Bruker

XFlash®6/60 electric refrigeration energy dispersive spectrometer (EDS). It offers an acceleration voltage range of 0.5–30 kV (0.1 kV step), magnification from 140× to 600,000×, and a 5-axis motorized stage (X: 0–100 mm, Y: 0–50 mm, Z: 3–65 mm, Tilt: -20° to +90°, Rotation: 360°). The EDS system has a 60 $mm^2$ detection area, an energy resolution better than 129 eV, and detects elements from Be (4) to Cf (98).

### 2.3 Spectroscopic Characterization Unit

#### 2.3.1. Horiba LabRAM HR Evolution Confocal Raman Microscope (see Fig. 1(h))

This fully integrated UV-Vis-NIR (220–2200 nm) confocal Raman system features an open-space microscope, hard-coupled for stability, to accommodate large sample holders (e.g., cryostats, pressure cells). It includes objectives (10×/0.25, 100×/0.9, LWDD 50×/0.5), 532 nm (100 mW) and 632.8 nm (17 mW) lasers with corresponding filters, and an achromatic spectrometer with a Raman shift range of 50–9000 $cm^{-1}$ (532 nm) and 50–6000 $cm^{-1}$ (632.8 nm). Detection is via a TE-cooled CCD (-70°C, 1024×256 pixels). The system is equipped for ultra-low frequency (ULF) and polarization measurements. An optional high-resolution helium flow cryostat enables temperature-dependent studies from 5 K to 450 K. A DAC high-pressure cell with a compatible temperature stage allows combined high-pressure (0–70 GPa) and variable-temperature (5–450 K) Raman experiments.

#### 2.3.2. Bruker VERTEX80V Fourier Transform Infrared Spectrometer (see Fig. 1(i))

The VERTEX80V covers a broad spectral range from 25,000 to 10 $cm^{-1}$ with a resolution better than 0.2 $cm^{-1}$. It is equipped with a suite of detectors: a liquid He-cooled Si bolometer (8–650 $cm^{-1}$), a FIR DLATGS detector (15–700 $cm^{-1}$), a MIR DLATGS detector (350–12,000 $cm^{-1}$), a liquid $N_2$-cooled high-sensitivity MCT detector (420–12,000 $cm^{-1}$), an InGaAs diode detector (4000–12,800 $cm^{-1}$), and a Si diode detector (9000–25,000 $cm^{-1}$). The system supports sample characterization over a temperature range of 5–350 K.

### 2.4 High-Quality Crystal Growth Unit

#### 2.4.1. RIBER COMPCT21 DZ Molecular Beam Epitaxy System (see Fig. 1(j))

This MBE system is designed for wafers up to 3 inches and achieves an ultimate base pressure of $3.8\times10^{-11}$ Torr. Source cells include valved cracker cells for As (VAC500) and Sb (VCOR300), high-capacity effusion cells for Ga and In (ABI85), a double-filament Al effusion cell (ABN60D), and a dopant cell for Si, Be, and GaTe (ABN135DC8). It is suitable for growing AlGaAs and InGaAs films with relevant doping. Key diagnostics include a beam flux gauge (Bayard-Alpert, $10^{-3}$–$10^{-10}$ Torr), a Hiden HALO 201 RC residual gas analyzer, and a 12 keV STAIB RHEED12 system. In collaboration with Peking University, an external cryocooler has been installed to facilitate the growth of high-quality Al-based superconducting thin-film heterostructures at lower temperatures.[6]

#### 2.4.2. High-Temperature Furnace Suite

This suite comprises over 20 furnaces for sintering, melting, and annealing under various atmospheres or vacuum. It includes general Muffle furnaces (up to 1700°C, ±1%), two-zone tube furnaces (up to 1200°C, ±1%, ~1 Pa vacuum), Bridgman furnaces (up to 1200°C, ±1%, ~1 Pa vacuum, growth speed 30–30,000 μm/hr), and induction furnaces (up to 2000°C, IR thermometer control). This equipment supports crystal growth via chemical vapor transport, flux, and Bridgman methods, as well as sample post-processing.

## 3. Representative Scientific Achievements Enabled by the F2 Station

The advanced instrumentation and expert support at the F2 Station have been instrumental in numerous high-impact research projects. Below are selected examples:

### 3.1 Discovery of Spin Supersolid and Giant Magnetocaloric Effect in $Na_2BaCo(PO_4)_2$ (see Fig. 3)

Exploring novel quantum states in frustrated magnets represents a major frontier in condensed matter physics, with promising implications for solid-state refrigeration. A collaborative effort by Researcher Peijie Sun and Associate Researcher Junsen Xiang (IOP CAS), Professor Gang Su (UCAS), Researcher Wei Li (ITP CAS), and Professor Wentao Jin (Beihang University) employed ultra-low-temperature thermodynamic measurements, neutron diffraction, and theoretical analysis to identify a spin supersolid state in the triangular lattice antiferromagnet $Na_2BaCo(PO_4)_2$. This phase simultaneously hosts diagonal long-range order (solid component) and off-diagonal quasi-long-range order (superfluid component). A giant magnetocaloric effect (MCE) is observed in the spin supersolid regime, underscoring its potential for low-temperature cooling applications. A prototype device demonstrates adiabatic demagnetization cooling from 2 K to below 94 mK. This behavior highlights a viable route toward helium-free refrigeration to sub-kelvin temperatures using solid-state materials.

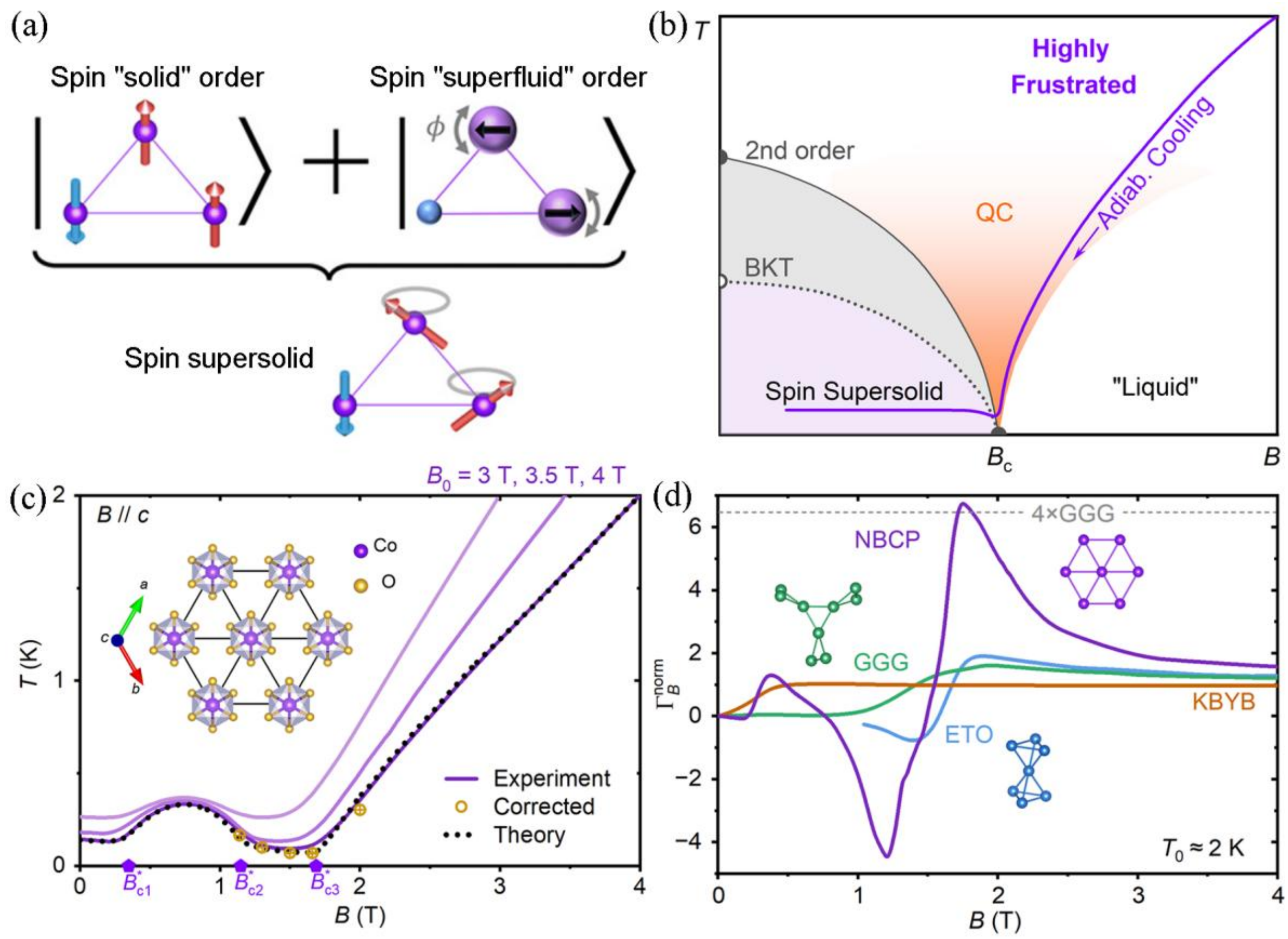

**Fig. 3. Schematic illustration of spin supersolid formation on a two-dimensional triangular lattice, and measured magnetocaloric effect of $Na_2BaCo(PO_4)_2$ during the demagnetization process. Figure (c, d) were obtained from the F2 station.**[1]

**F2 Station Contribution:** The majority of the experimental work in this study was carried out at the F2 station. Specifically, F2 station engineers worked in close collaboration with the user team to develop the low-temperature quasi-adiabatic magnetocaloric measurement setup (described in Section 2.1.1 and shown in Fig. 2(a)), a capability that was essential for detecting the characteristic plateau features indicative of the spin supersolid phase. The critical MCE data thus obtained, showing two distinct adiabatic demagnetization plateaus that provide direct thermodynamic evidence for the coexistence of spin solid order and spin superfluidity, were published as main-text figures (corresponding to Fig. 2c and Fig. 3 of the original publication).

Without the F2 station's unique combination of instrument infrastructure and specialized technical expertise in ultra-low-temperature calorimetry, these decisive measurements could not have been accomplished.

**Publication:** Nature 625, 270–275 (2024); selected as one of the "Top 10 Scientific Advances in China for 2024" and was also recognized among the "Top 10 Scientific and Technological Progress News" by academicians of the Chinese Academy of Sciences and the Chinese Academy of Engineering.

### 3.2 Pressure-Induced Bulk Superconductivity in Trilayer Nickelate $La_4Ni_3O_{10-\delta}$ (see Fig. 4)

Following the discovery of high-$T_c$ superconductivity in bilayer $La_3Ni_2O_7$, the existence of bulk superconductivity and other superconducting nickelate phases remained open questions. A collaborative team led by Prof. Jun Zhao (Fudan University), Researcher Jiangang Guo (IOP CAS), and Researcher Qiaoshi Zeng (HPSTAR) applied high pressure to trilayer $La_4Ni_3O_{10-\delta}$ single crystals. At 69.0 GPa, a superconducting transition at $T_c \approx 30$ K was observed. DC magnetometry confirmed a diamagnetic shielding fraction >80%, providing definitive evidence for bulk superconductivity. The simultaneous observation of strange metal behavior (linear-in-$T$ resistivity up to 300 K) offered new insights into high-$T_c$ mechanisms. The distinct chemical environments of the inner and outer $NiO_2$ layers in the trilayer structure provide a unique platform for studying the roles of interlayer coupling and charge transfer.

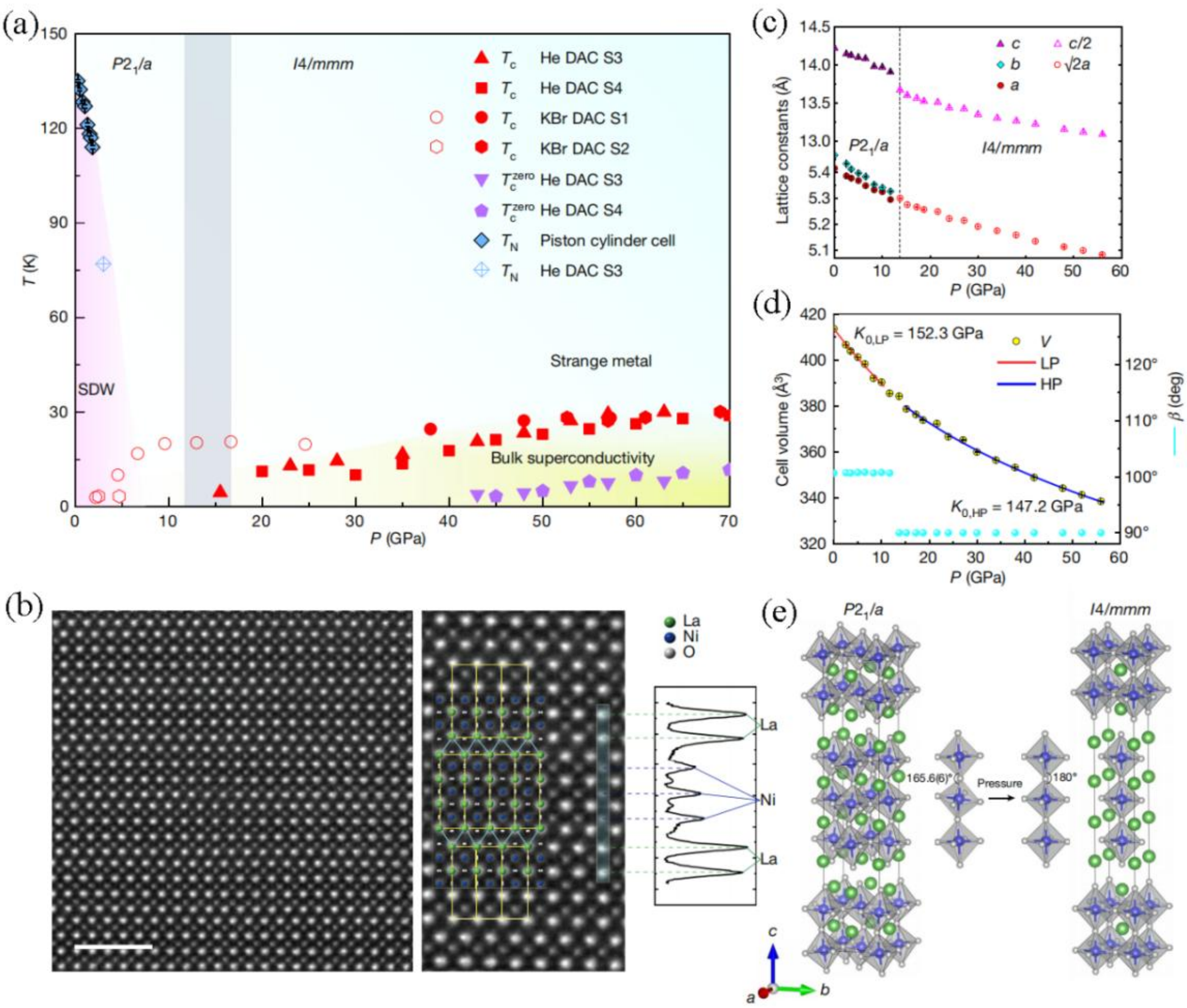

**Fig. 4. Pressure-dependent phase diagram and lattice structure of $La_4Ni_3O_{10-\delta}$. Part of the $T_c$ data in Figure (a) were obtained from the F2 station.**[7]

**F2 Station Contribution:** A large number of the experiments in this work were completed at the F2 station. Senior Engineer Xu Chen carried out extensive high-pressure electrical transport experiments using the self-built DAC-based in-situ high-pressure system integrated with the station's PPMS (described in Section 2.1.1). Specifically, he performed systematic preliminary high-pressure screening, comprehensive property characterization, and rigorous phase identification on multiple batches of user-supplied $La_4Ni_3O_{10-\delta}$ single crystals. These methodical efforts led to the first successful acquisition of high-quality superconducting transition signals (zero-resistance state) and linear-in-temperature resistivity signatures (strange metal behavior) under pressures up to 69 GPa. These data provided the pivotal experimental evidence establishing that $La_4Ni_3O_{10-\delta}$ is a bulk superconductor. The corresponding data were published as main-text figures (corresponding to Fig. 1a and Fig. 3e of the original publication).

**Publication:** Nature 631, 531–536 (2024); featured in Nature "News & Views".

### 3.3 Revealing Discrete Degeneracy in Metallic Kagome Spin Ice via Anomalous Hall Effect (see Fig. 5)

Geometrically frustrated metals offer platforms to study interplay between spin ice rules and itinerant electrons. An international team led by Professor Kan Zhao (IOP CAS), Professor Hua Chen (Colorado State University), and Professor Philipp Gegenwart (University of Augsburg) conducted detailed magnetotransport studies on high-quality HoAgGe single crystals, which host a kagome spin ice state. Hall and magnetoresistance measurements at the 1/3 and 2/3 magnetization plateaus revealed clear hysteresis not present in the magnetization, indicating the existence of at least two degenerate states with identical magnetization but different transport properties—a time-reversal-like degeneracy. This demonstrated the unique power of the anomalous Hall effect in probing hidden symmetries in frustrated metals.

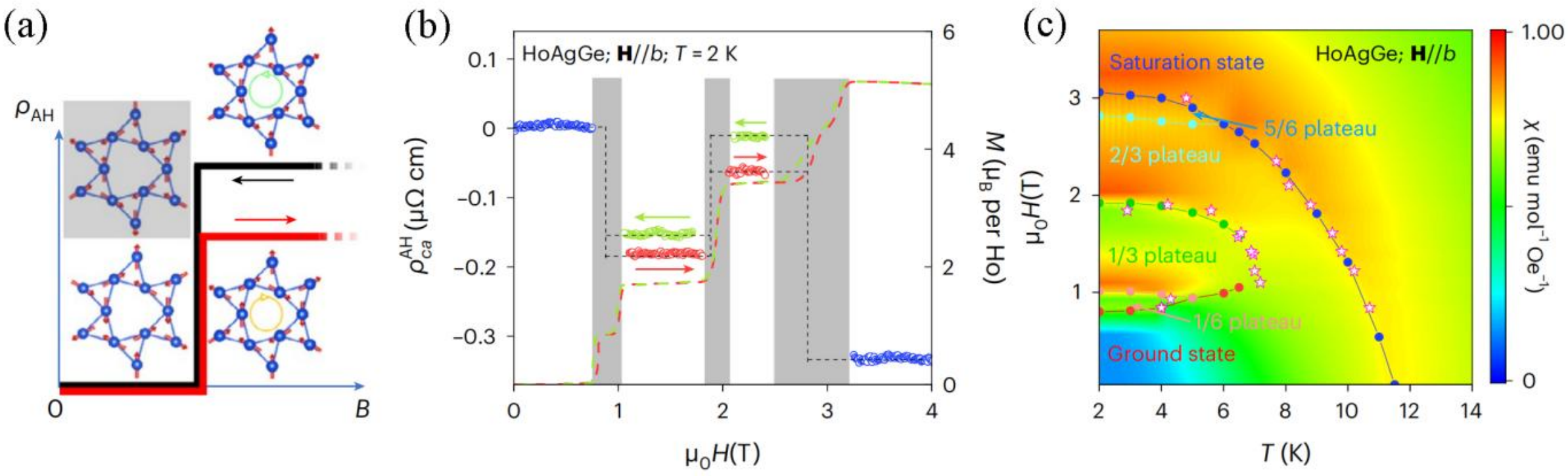

**Fig. 5. Time-reversal-like degeneracies in field-induced plateau phases of HoAgGe. The anomalous Hall resistivity data were obtained from the F2 station.**[8]

**F2 Station Contribution:** The station conducted a large number of low-temperature transport tests, providing the high-quality Hall and magnetoresistance data. These measurements confirmed the realization of a kagome spin ice state in the frustrated metal HoAgGe. The corresponding data were published as main-text figures (corresponding to Fig. 6 of the original publication).

**Publication:** Nature Physics 20, 442–449 (2024).

**3.4 Hydrogen-Bond-Controlled Metal-Insulator Transition with $10^7$ Resistance Change (see Fig. 6)**

Hydrogen bonds typically do not drastically alter electronic conductivity. A team led by Distinguished Researcher Tianping Ying, Researcher Jiangang Guo, and Researcher Xiaolong Chen (IOP CAS), with Researcher Yurui Gao (NCNST), inserted 1,3-diaminopropane (DAP) into $SnSe_2$. The dynamic disorder ("breaking" and "bonding") of the hydrogen bonds above 160 K becomes partially frozen below 160 K. This "dynamic-to-static" transition drives a metal-insulator transition with a resistivity change spanning seven orders of magnitude. S substitution for Se strengthens the interaction, raising the transition temperature toward room temperature.

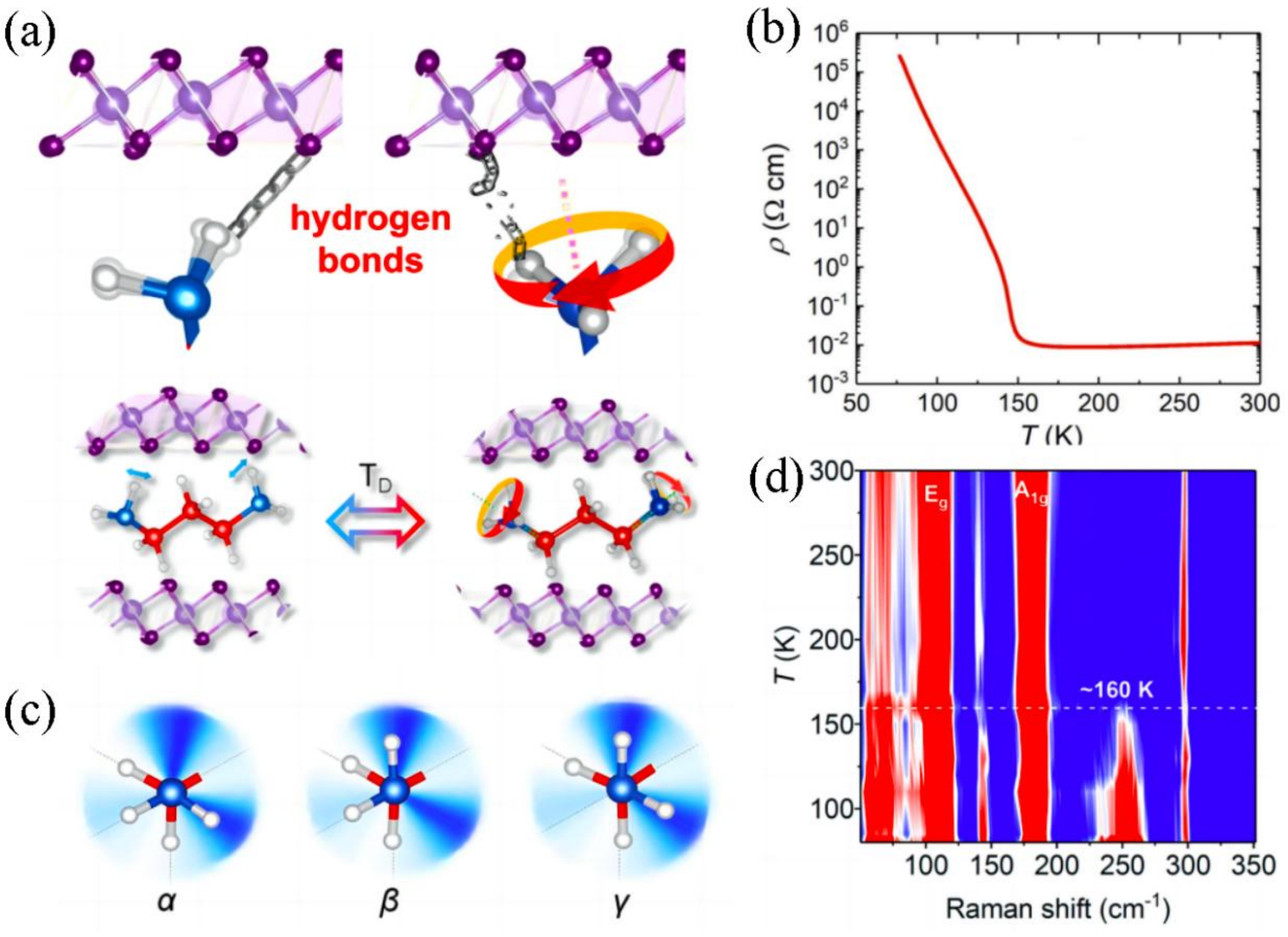

**Fig. 6. Schematic illustration of the 1,3-DAP molecular states in $(1,3\text{-DAP})_{0.5}SnSe_2$ at different temperatures, accompanied by temperature-dependent electrical resistivity and Raman spectra.**[9]

**F2 Station Contribution:** The station grew the high-quality $SnSe_2$ single crystals via chemical vapor transport, providing the essential material foundation for the subsequent achievement of a metal-insulator transition with a resistivity change spanning seven orders of magnitude.

**Publication:** Nature Chemistry 16, 1803–1810 (2024).

### 3.5 Giant Anomalous Hall Angle and Magnetic Weyl Hall Sensing (see Fig. 7)

The anomalous Hall angle, crucial for applications, has historically been small (<5%). Enke Liu's team (IOP CAS) developed a two-variable model for the anomalous Hall angle and demonstrated it in the magnetic Weyl semimetal $Co_3Sn_2S_2$. By tuning longitudinal resistivity and anomalous Hall conductivity via temperature, thickness, and doping, they achieved a giant zero-field anomalous Hall angle of 25° (46%). They fabricated an anomalous Hall sensor with a sensitivity of 7028 μΩ·cm/T and a magnetic field detectability of 23 nT/√Hz at 1 Hz, outperforming previous devices.

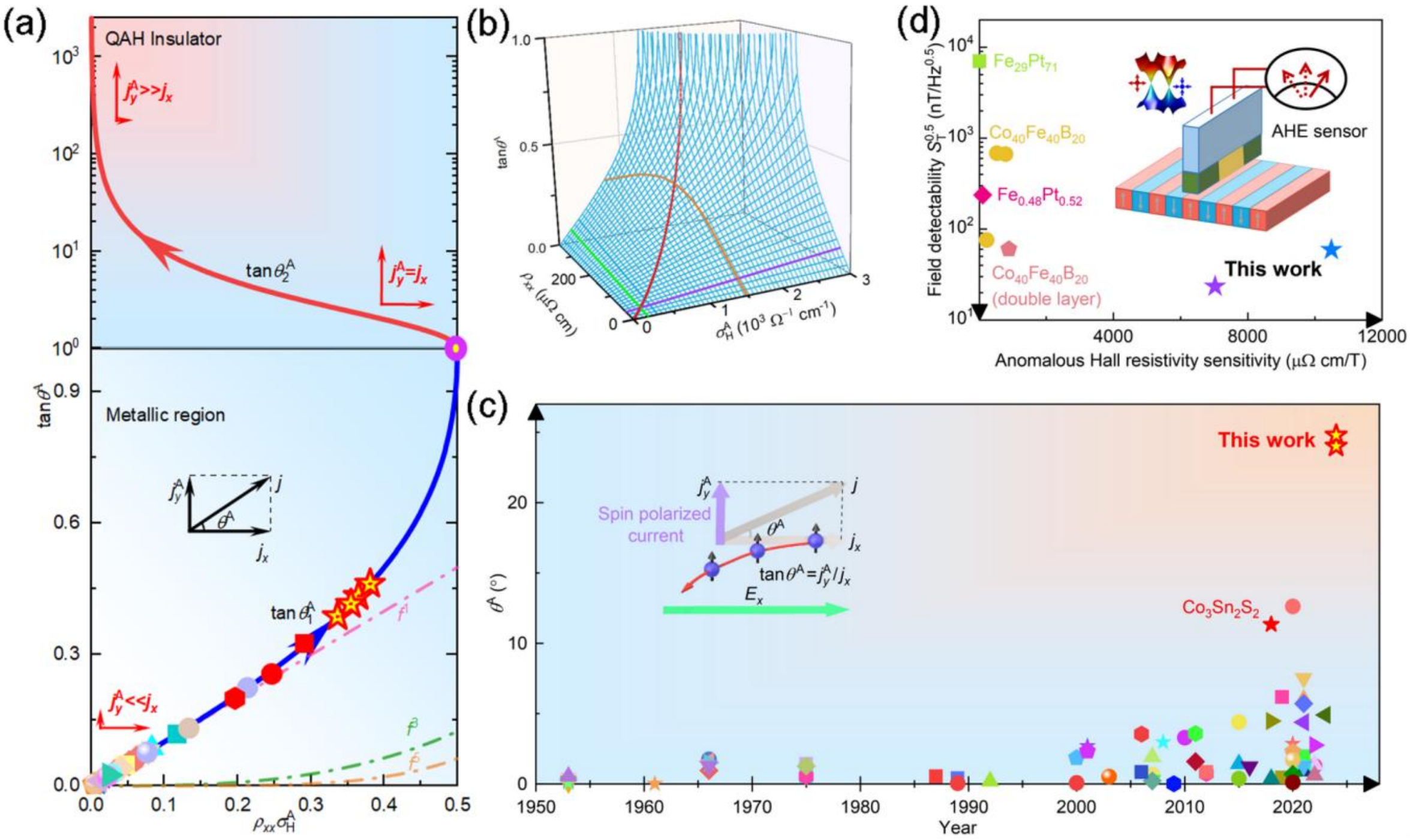

**Fig. 7. Two-variable functional relationship of the anomalous Hall angle, the giant anomalous Hall angle, and the anomalous Hall sensing based on magnetic Weyl semimetals.**[10]

**F2 Station Contribution:** The station conducted extensive preliminary low-temperature transport experiments, laying a solid foundation for the smooth progress of this work.

**Publication:** Nature Electronics 8, 386–393 (2025).

### 3.6 Discovery of Hypergap Transparent Conductors (see Fig. 8)

Intrinsic transparent metals, proposed decades ago, had never been realized. Researcher Ling Lu's team (IOP CAS) predicted and experimentally discovered such an electronic state in organic charge-transfer salts $(TMTTF)_2X$. The "hypergap"—an absorption-free window between intraband and interband transitions—arises from an isolated, narrow metallic band. Bulk $(TMTTF)_2X$ crystals showed significant transparency from visible red to near-infrared (up to 30 μm thickness), with a minimum dielectric loss (~0.01) comparable to ITO.

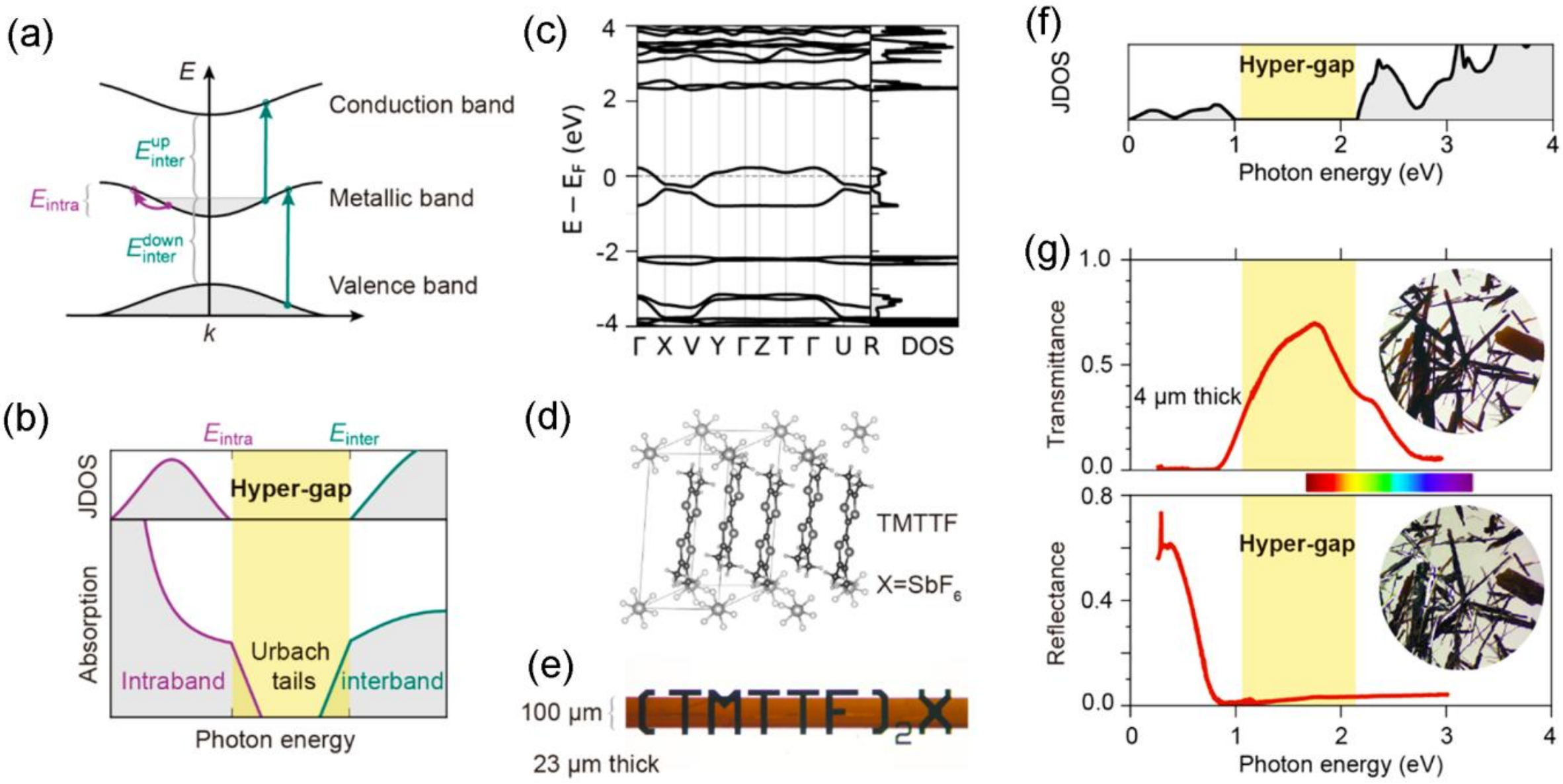

**Fig. 8. Theoretical principle, materials prediction, and experimental discovery of hypergap transparent conductors.**[11]

**F2 Station Contribution:** The station conducted preliminary transport characterization experiments, creating the necessary conditions for the users to better understand the fundamental physical properties of these transparent conductors.

**Publication:** Nature Materials 24, 1387–1392 (2025) (Cover Story); selected as one of the "Top Ten Optical Advances in China for 2025".

### 3.7 Evidence of Molecular Water in Chang'e-5 Lunar Soil (see Fig. 9)

The existence and form of water on the Moon are fundamental questions. A team led by Researcher Xiaolong Chen and Researcher Shifeng Jin (IOP CAS) discovered a previously unknown hydrated mineral, ULM-1 ($NH_4MgCl_3 \cdot 6H_2O$), in Chang'e-5 samples. Single-crystal XRD, IR/Raman spectroscopy, and chemical analysis confirmed the presence of molecular water (up to 41 wt.%) and ammonium, providing direct evidence for stable hydrated salts on the Moon and placing constraints on lunar volcanic gas composition.

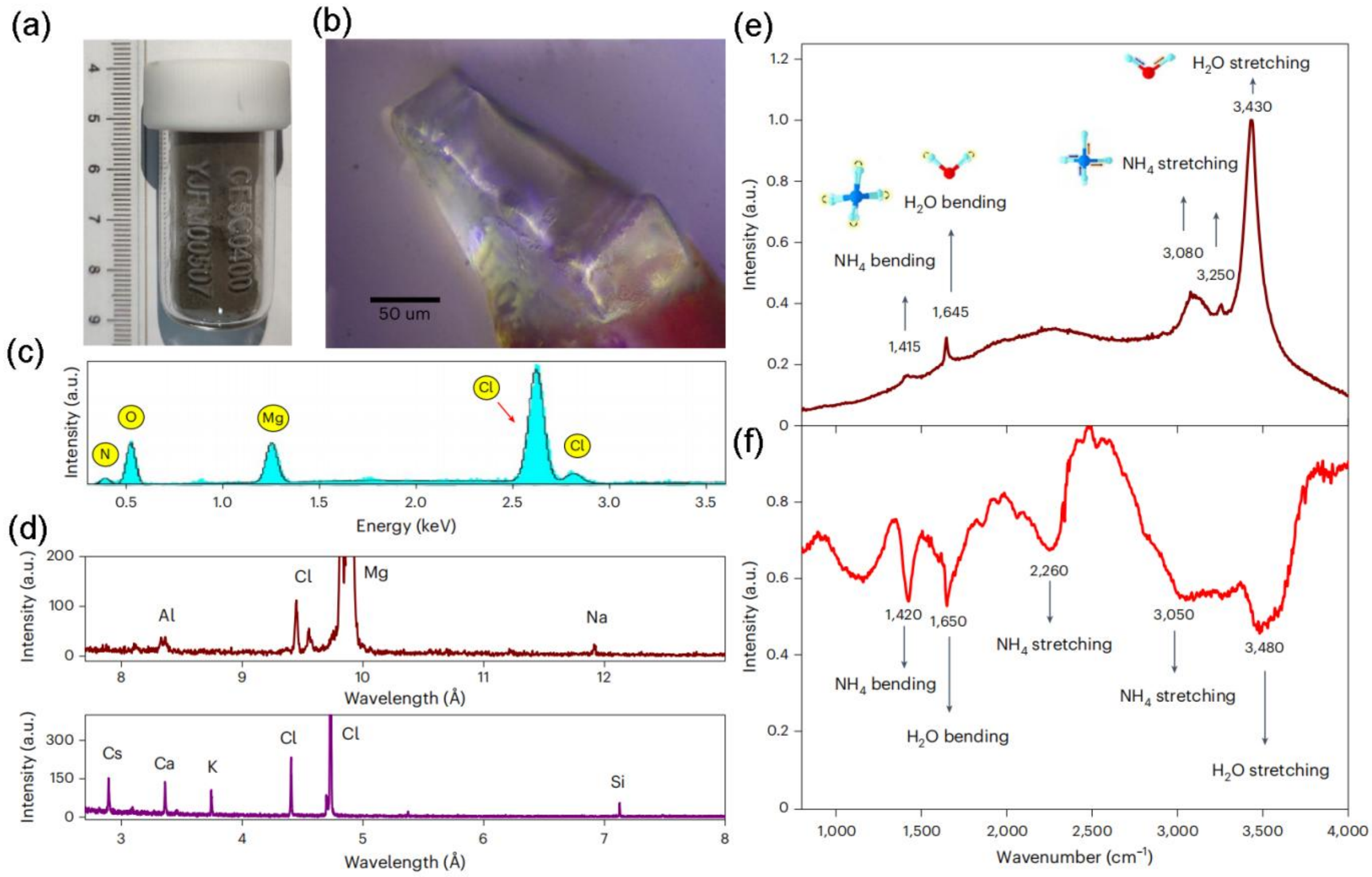

**Fig. 9. Photograph and composition of ULM-1. The photograph of ULM-1 in Figure (b) was obtained from the F2 station.**[12]

**F2 Station Contribution:** Conducted extensive single-crystal XRD screening of lunar soil samples and obtained high-resolution, lossless sample photographs. The data were published as main-text figures (corresponding to Fig. 1b of the original publication).

**Publication:** Nature Astronomy 8, 1127–1137 (2024).

## 4. Conclusion and Outlook

The Sample Pre-selection and Characterization Station at SECUF has established itself as a vital infrastructure node, providing integrated services across sample growth, processing, screening, and comprehensive characterization. Its well-equipped units and expert staff have successfully supported a wide user community. To date, the station has accumulated over 100,000 hours of user operation, served more than 373 research projects, and supported users from over 60 external institutions. User work facilitated by the F2 Station has resulted in over 230 acknowledged publications, including papers in Nature, Nature sister journals, National Science Review, Physical Review X/Letters, and JACS, as well as more than 30 patents. One acknowledged achievement was selected as China's Top 10 Scientific Advances (2024), seven were selected as Top Ten Highlights of SECUF in 2024, one acknowledged achievement was selected as Top Ten Optical Advances in China for 2025, and three were selected as Top Ten Highlights of SECUF in 2025.

Looking forward, the F2 Station will continue to enhance its technical capabilities through the development of new integrated measurement systems (e.g., combined high-pressure, low-temperature, and magnetic field environments) and deeper collaboration with users to tackle cutting-edge scientific challenges. We welcome researchers to utilize the facilities at the F2 Station, engage with our technical staff, and collaborate to advance cutting-edge research in condensed matter and materials science.

**Acknowledgments**

The authors thank the staff and users of the SECUF F2 Station for their contributions. We also acknowledge Prof. P. J. Sun, Prof. J. S. Xiang, Prof. G. Su, Prof. W. Li, Prof. W. T. Jin, Prof. J. Zhao, Prof. J. G. Guo, Prof. Q. S. Zeng, Prof. B. Y. Pan, Prof. K. Zhao, Prof. H. Chen, Prof. P. Gegenwart, Prof. X. L. Chen, Prof. T. P. Ying, Prof. Y. R. Gao, Prof. E. K. Liu, Prof. L. Lu, and Prof. S. F. Jin, for their contributions to the studies listed in this manuscript.

**References:**

[1] Xiang J, Zhang C, Gao Y, Schmidt W, Schmalzl K, Wang C W, Li B, Xi N, Liu X Y, Jin H, Li G, Shen J, Chen Z, Qi Y, Wan Y, Jin W, Li W, Sun P and Su G 2024 *Nature* **625** 270
[2] 陈旭,任会芬,孙涛,崔敏杰,边勇波,杨明章,王俊杰,苏少奎 用于磁性测量系统氦-3温区的金刚石对顶压砧装置及其使用方法 202610298225.2
[3] 陈旭,任会芬,孙涛,崔敏杰,边勇波,苏少奎 用于磁性测量系统氦-3温区的双螺旋活塞加压装置及其使用方法 202610410271.7
[4] 陈小龙，柴聪聪，盛达，陈旭，金士锋，宋友庭，孙涛用于原位调控晶体外加电场的单晶衍射样品台装置 202410117292.0
[5] 姜小明,郭国聪 电子结构晶体学 9787030723536
[6] 张文锋,姚嘉宁,杜瑞瑞,苏少奎 应用于分子束外延设备的可位移冷却样品台 202521147530.9
[7] Zhu Y, Peng D, Zhang E, Pan B, Chen X, Chen L, Ren H, Liu F, Hao Y, Li N, Xing Z, Lan F, Han J, Wang J, Jia D, Wo H, Gu Y, Gu Y, Ji L, Wang W, Gou H, Shen Y, Ying T, Chen X, Yang W, Cao H, Zheng C, Zeng Q, Guo J G and Zhao J 2024 *Nature* **631** 531
[8] Zhao K, Tokiwa Y, Chen H and Gegenwart P 2014 *Nat. Phys.* **20** 442
[9] Xie Z, Luo R, Ying T, Gao Y, Song B, Yu T, Chen X, Hao M, Chai C, Yan J, Huang Z, Chen Z, Du L, Zhu C, Guo J and Chen X 2024 *Nat. Chem.* **16** 1803
[10] Yang J, Shang Y, Liu X, Wang Y, Dong X, Zeng Q, Lyu M, Zhang S, Liu Y, Wang B, Wei H, Wu Y, Parkin S, Liu G, Felser C, Liu E and Shen B 2025 *Nat. Electron.* **8** 386
[11] Wu Z, Li C, Hu X, Chen K, Guo X, Li Y and Lu L 2025 *Nat. Mater.* **24** 1387
[12] Jin S, Hao M, Guo Z, Yin B, Ma Y, Deng L, Chen X, Song Y, Cao C, Cai C, Wei Q, Ma Y, Guo J and Chen X 2024 *Nat. Astron.* **8** 1127